\documentclass[
  aps,
  pre,
  reprint,
  superscriptaddress,
  longbibliography
]{revtex4-2}

\usepackage{amsmath}
\usepackage{amssymb}
\usepackage{graphicx}
\usepackage{booktabs}
\usepackage{dcolumn}
\usepackage{bm}
\usepackage{xcolor}
\usepackage{hyperref}

\definecolor{RevisionBlue}{RGB}{0,91,150}

\definecolor{RevisionGreen}{HTML}{2E7D32}

\hypersetup{
    colorlinks=true,
    linkcolor=blue,
    citecolor=blue,
    urlcolor=blue
}

\begin{document}

\title{Unambiguous characterization of in-plane dielectric response in nanoconfined liquids: Water as a case study}

\author{Jon Zubeltzu}
\email{jon.zubeltzu@ehu.eus}
\affiliation{Donostia International Physics Center (DIPC), Manuel Lardizabal Ibilbidea 4, 20018 Donostia, Spain}
\affiliation{Department of Applied Physics, Engineering School of Gipuzkoa, Basque Country University, UPV/EHU, Europa Plaza 1, 20018 Donostia, Spain}

\begin{abstract}
The in-plane dielectric constant of nanoconfined water has attracted growing interest in recent years. Nevertheless, assigning a bulk-like dielectric constant to a molecularly thin film is convention dependent because it requires choosing an effective water width. We propose the in-plane 2D polarizability, $\alpha_{\parallel}$, as a macroscopic characterization of the in-plane dielectric response that avoids this choice. Using classical molecular dynamics simulations, we compute $\alpha_{\parallel}$ via two independent and consistent methods: based on fluctuation-dissipation theory, and from the induced dipole moment when water is placed in a capacitor.
Our results provide the framework to quantify the in-plane dielectric response of polar liquids across simulations and experiments.
\end{abstract}

\maketitle
\section{Introduction}
\label{sec:intro}

Polar fluids and electrolyte solutions are central to many natural and industrial processes. Under extreme confinement in nanometer-scale pores, their behavior can depart qualitatively from the bulk, exhibiting anomalous transport, shifted phase boundaries, strong correlation effects, and dielectric anomalies~\cite{aluru2023}. Water is a natural focus within this broader class: as the prototypical polar liquid, it is ubiquitous in interfacial and nanoscale environments and therefore serves as a key system for studying confinement-induced effects.

A central aspect of these confinement-induced changes is the dielectric response. Before direct experimental benchmarks were available, computational simulations had already indicated that the dielectric response of interfacial water differs substantially from that of the bulk~\cite{ballenegger2005,marti2006,zhang2013,zubeltzu2016,schlaich2016,de2016}. A consistent finding of these studies is a pronounced dielectric anisotropy, with distinct in-plane and out-of-plane components. Experimentally, Fumagalli \textit{et al.} reported a strong reduction of the perpendicular dielectric constant for water confined in slabs with thicknesses on the order of one nanometer~\cite{fumagalli2018}. This finding triggered extensive theoretical and computational efforts to explain the origin of the apparent low permittivity. Some studies attribute it to intrinsically structured interfacial layers that behave as a low-permittivity “dead layer”~\cite{dufils2024,papadopoulou2021,ahmadabadi2021,zhang2013}, whereas others emphasize long-ranged collective dipolar correlations~\cite{olivieri2021,monet2021,mondal2021}.

More recently, experiments in slit like nanochannels reported that the in-plane permittivity of few-layer water becomes extremely large, accompanied by a pronounced increase in conductivity. These observations were interpreted as signatures of an unusually polarizable, ferroelectric-like state associated with hydrogen-bond disorder and proton dynamics~\cite{Wang2025}. In line with these results, several studies based on molecular dynamics had found already that the in-plane dielectric permittivity can increase dramatically relative to bulk under extreme, few-layer confinement~\cite{motevaselian2020,renou2015,hamid2021}, anticipating and qualitatively reproducing the experimentally reported enhancement.

Interpreting in-plane permittivities quantitatively at the nanoscale, however, is not straightforward: assigning a dielectric constant to a molecularly thin water film is convention dependent because it requires choosing a thickness (or width) $w$ for the water layer. This thickness is not uniquely defined at the nanoscale, and the resulting permittivity is highly sensitive to the particular choice of $w$. Recent theoretical analyses have therefore argued that, under molecular-scale confinement, the appropriate observable is instead a two-dimensional (2D) polarizability, and have determined this quantity for the perpendicular response of nanoconfined water~\cite{zubeltzu2025}. The same approach had already been adopted for atomically thin crystals (e.g., graphene), where the thickness of such materials is ill-defined and the dielectric response is naturally expressed in terms of a 2D polarizability rather than a three-dimensional (3D) permittivity~\cite{cudazzo2011,tian2019,olsen2016}.

In light of these considerations, here we adopt a 2D characterization of the in-plane dielectric response, using slit-confined water as a case study. In close analogy with recent 2D response measures introduced for the perpendicular component~\cite{zubeltzu2025}, we characterize the in-plane response of nanoconfined fluids through the in-plane 2D polarizability, a 2D response coefficient that corresponds to the in-plane dielectric profile integrated across the film~\cite{kornyshev1981,bonthuis2012,stark2025static} and whose definition does not involve an effective film width. Rather than focusing on reproducing the experimentally reported giant in-plane permittivities—an outcome already addressed by simulations~\cite{motevaselian2020,renou2015,hamid2021}—we instead establish a practical protocol to quantify the in-plane dielectric response at the nanoscale. We compute the in-plane 2D polarizability of slit-confined water using classical molecular dynamics (MD) simulations via two independent, mutually consistent routes: (1) a fluctuation–dissipation estimate based on polarization fluctuations in the absence of an applied electric field, and (2) a field-response estimate obtained from the induced dipole moment when the confined water is placed in a capacitor held at fixed potential difference. The latter method shows that $\alpha_{\parallel}$ can, in principle, be determined directly from experiments through measurements of the induced charge on the capacitor plates, whereas the former is computationally far more efficient and avoids the finite-size boundary effects inherent to the capacitor geometry. As an additional consistency check, we evaluate the perpendicular 2D polarizability from the corresponding fluctuation-dissipation expression and find excellent agreement with previously reported values obtained from field-response calculations~\cite{zubeltzu2025}. Taken together, the results of this work and those of Ref.~\cite{zubeltzu2025} provide a unified framework to quantify and compare the dielectric response of nanoconfined polar liquids, both in-plane and perpendicular, across simulations and experiments.

\section{Theoretical framework}

In macroscopic electrostatics and within linear response, the dielectric constant is introduced through the relation between the polarization and the applied electric field,
\begin{equation}
\mathbf{P}=\varepsilon_0(\varepsilon-1)\mathbf{E},
\end{equation}
where $\mathbf{P}$ is the three-dimensional polarization density (dipole moment per unit volume) and $\varepsilon$ is the dielectric constant. This formulation is inherently bulk-like, since it requires a well-defined three-dimensional polarization density and therefore an unambiguous volume over which the dipole moment is normalized. For quasi-2D systems, such as atomically thin crystals~\cite{tian2019} or, as analyzed in this work, nanoconfined liquids in the few-layer limit, this requirement becomes problematic because it requires assigning an effective thickness $w$, for which no unique definition exists. In the experimental work of Fumagalli \textit{et al.}, based on AFM nanocapacitance measurements~\cite{fumagalli2018}, converting the measured capacitance gradient $dC/dz$ into a dielectric constant requires modeling the system geometry, including the dielectric thickness~\cite{fumagalli2018,stark2025static}. As discussed in~\cite{zubeltzu2025,stark2025static,dufils2024}, different reasonable choices of $w$ can lead to markedly different values of $\varepsilon$.

A natural way to avoid the thickness ambiguity at molecular scales is to formulate the dielectric response in strictly 2D form, using the 2D polarization $P_{2\mathrm{D}}$ (dipole moment per unit area), which remains well defined in the few-layer regime. This viewpoint has been adopted for atomically thin crystals, where the dielectric response can be expressed in terms of 2D polarizabilities for both parallel and perpendicular directions~\cite{tian2019}. For nanoconfined liquids, an analogous 2D formulation has recently been developed for the perpendicular component~\cite{zubeltzu2025}; here we extend the same approach to the in-plane response. In a homogeneous planar system with invariance along the $x$ and $y$ directions, Maxwell’s equation $\nabla\times\mathbf{E}(z)=0$ implies that the parallel component of the electric field is continuous, so $E_{\parallel}$ is uniform. We therefore characterize in-plane screening by the in-plane 2D polarizability, defined as the linear-response coefficient relating the induced 2D polarization to the in-plane electric field,
\begin{equation}
\alpha_{\parallel}\equiv \frac{1}{\varepsilon_0}\frac{\partial P_{2\mathrm{D}}}{\partial E_{\parallel}}.
\label{def}
\end{equation}
This quantity is well defined provided that the in-plane area of the system is well defined. As in the case of $\alpha_{\perp}$, $\alpha_{\parallel}$ is intensive with respect to the in-plane dimensions and extensive along the confining direction: it depends on the physical amount of confined liquid, for example on the number of molecular layers.

For a planar system with in-plane translational invariance, the nonlocal dielectric response reduces to position-dependent profiles $\varepsilon_{\parallel}(z)$ and $\varepsilon_{\perp}(z)$~\cite{stark2025static}. Under a uniform in-plane field, $P_{\parallel}(z)=\varepsilon_0\left[\varepsilon_{\parallel}(z)-1\right]E_{\parallel}$, and integrating across the system gives
\begin{equation}
\alpha_{\parallel}=\int\left[\varepsilon_{\parallel}(z)-1\right]dz .
\label{eq:alpha_par_profile}
\end{equation}
Analogously, for a uniform displacement field $D_{\perp}$, $P_{\perp}(z)=D_{\perp}\left[1-1/\varepsilon_{\perp}(z)\right]$ and
\begin{equation}
\alpha_{\perp}=\int\left[1-\frac{1}{\varepsilon_{\perp}(z)}\right]dz .
\label{eq:alpha_perp_profile}
\end{equation}
The integrals extend over the whole system along the confinement direction, so no film width enters these definitions. Integrals of the dielectric profiles of this kind have been used previously, see for example Refs.~\cite{kornyshev1981,bonthuis2012}.

The physical meaning of $\alpha_{\parallel}$ is that of an in-plane characteristic electrostatic length. More specifically, following Tian \textit{et al.}~\cite{tian2019}, $\alpha_{\parallel}$ defines an in-plane screening radius that governs the in-plane crossover in the interaction produced by a point charge embedded in the film: at short distances the interaction is strongly affected by the dielectric response of the confined layer, whereas at large separations it gradually approaches the asymptotic vacuum Coulomb form. Thus, $\alpha_{\parallel}$ provides the intrinsic in-plane length scale associated with this gradual crossover. This interpretation is complementary to that of the perpendicular 2D polarizability, which can be viewed as an effective dielectric thickness~\cite{tian2019,zubeltzu2025}. Taken together, $\alpha_{\parallel}$ and $\alpha_{\perp}$ provide an anisotropic geometric representation of the dielectric response of the confined fluid, analogous to a flattened polarizability ellipsoid~\cite{tian2019}.

If an in-plane dielectric constant is nevertheless to be reported, it follows from Eq.~(\ref{eq:alpha_par_profile}) that the effective permittivity of a film of assigned width $w$ is~\cite{bonthuis2012,tian2019}
\begin{equation}
\varepsilon_{\parallel}=1+\frac{\alpha_{\parallel}}{w},
\label{definition_alpha}
\end{equation}
so that $\varepsilon_{\parallel}$ depends on the convention adopted for $w$, whereas $\alpha_{\parallel}$ does not.

Taking the 2D polarizability as the central response parameter, we determine both $\alpha_{\parallel}$ and $\alpha_{\perp}$ directly from equilibrium polarization fluctuations. Building on Kirkwood-Fr\"ohlich theory~\cite{kirkwood1939,Frohlich1958}, the dielectric response of planar systems can be estimated from polarization fluctuations using expressions that depend on the electrostatic boundary conditions~\cite{zhang2016computing,stark2025static}. In the slab geometries considered here, the in-plane component is associated with the response to a uniform $E_{\parallel}$, whereas the perpendicular component is associated with the response to a uniform $D_{\perp}$. Rewriting the corresponding fluctuation expressions for $\varepsilon_{\parallel}$ and $\varepsilon_{\perp}$~\cite{stark2025static,schlaich2016,loche2020universal,coelho2024interplay,ballenegger2005,motevaselian2020confinement,zhang2013,motevaselian2020universal,zhang2016computing} in terms of the 2D polarizabilities~\cite{tian2019}, and using effective 2D long-range electrostatics for the periodic slab~\cite{yeh1999}, we obtain
\begin{equation}
\alpha_{\parallel}
=\frac{\langle M_{\parallel}^2\rangle-\langle M_{\parallel}\rangle^2}{2\,k_{\mathrm{B}}T\,\varepsilon_0\,A}
\label{eq:alpha_parallel_fluc}
\end{equation}
and
\begin{equation}
\alpha_{\perp}
=\frac{\langle M_{\perp}^2\rangle-\langle M_{\perp}\rangle^2}{k_{\mathrm{B}}T\,\varepsilon_0\,A}.
\label{eq:alpha_perp_fluc}
\end{equation}
Here, $M_{\parallel}$ and $M_{\perp}$ denote, respectively, the parallel and perpendicular components of the total dipole moment of the system, $\langle\cdot\rangle$ indicates an ensemble average, $k_{\mathrm{B}}$ is the Boltzmann constant, $T$ is the temperature, $\varepsilon_0$ is the vacuum permittivity, and $A$ is the area of the dielectric. In contrast to the conventional fluctuation expressions for the dielectric-constant components, Eqs.~(\ref{eq:alpha_parallel_fluc})--(\ref{eq:alpha_perp_fluc}) involve only the area of the system and do not require assigning an effective width or volume to the confined film.

\section{In-plane 2D polarizability}

\subsection{Computational methods}

All MD simulations were performed with the LAMMPS package~\cite{LAMMPS}. Water was modeled using the SPC/E force field at $T=300$ K and confined along the $z$ direction by two Lennard-Jones 9-3 potentials, with their asymptotes separated by $8$~\AA. The parameters of the water-wall interaction were chosen to reproduce the interaction of water with solid paraffin, as in~\cite{zubeltzu2016}. Simulations were carried out in the $NVT$ ensemble using a Nosé-Hoover thermostat, and the equations of motion were integrated with a timestep of $1$~fs. Periodic-image interactions along the confinement direction were effectively canceled using the LAMMPS \texttt{slab} correction~\cite{yeh1999} with \texttt{volfactor} $=3.0$. For each value of the lateral supercell dimensions in the $x$ and $y$ directions, the number of water molecules was adjusted to keep a 2D molecular density of $\sigma=0.177$~\AA$^{-2}$. Along the $x$ direction, periodic boundary conditions (PBCs) were applied.

Two computational setups were employed, differing in the boundary conditions along the $y$ direction; in both cases, the electrostatic boundary conditions along the $x$ and $y$ directions correspond to a constant electric-field ensemble~\cite{zhang2016computing}:

\begin{figure}[t]
  \centering
  \includegraphics[width=1.02\columnwidth]{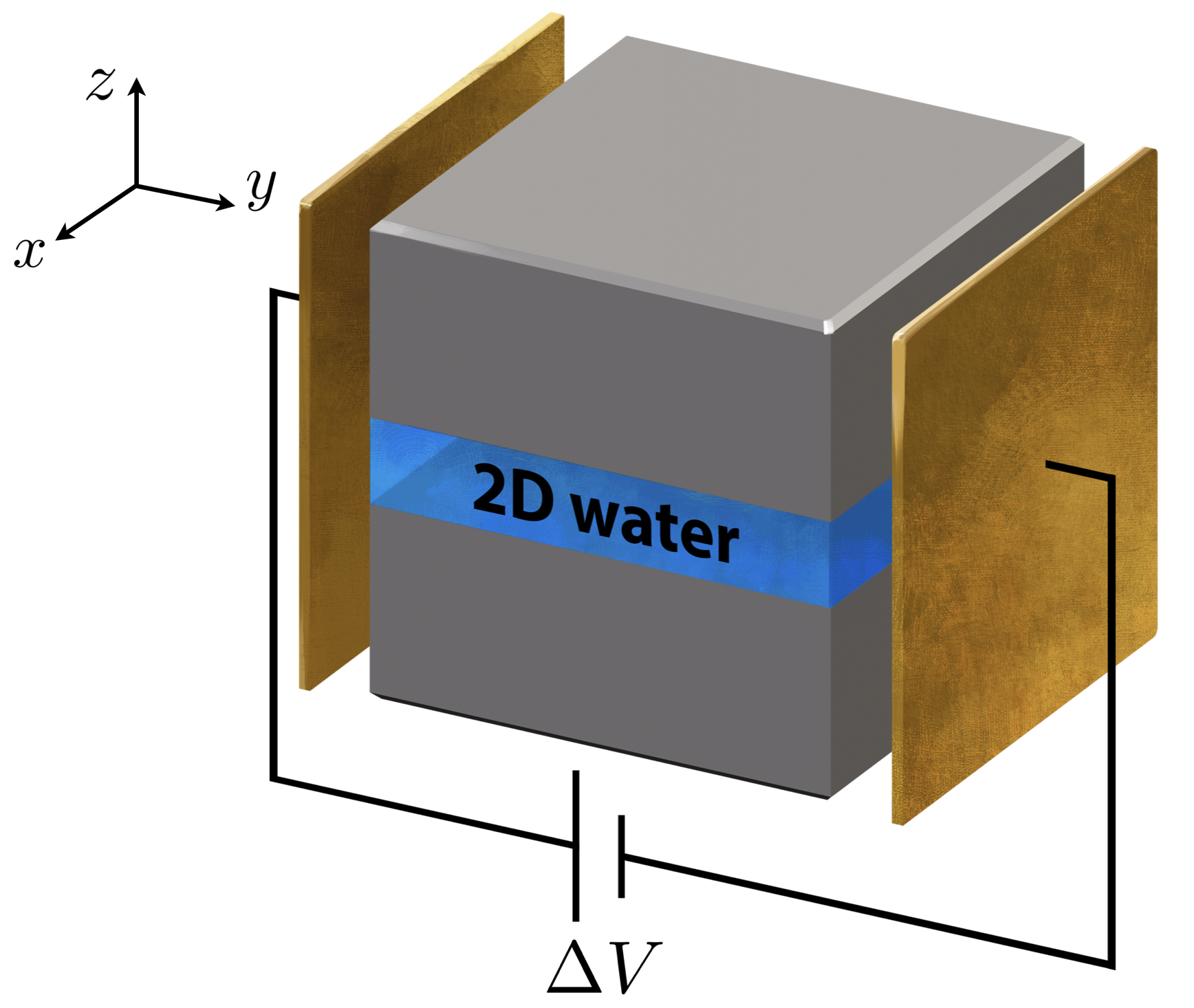}
  \caption{Schematic representation of the capacitor setup. Single-atom-thick Au electrodes of area $A=L_x\times L_z$ are located at $y=0$ and $y=l_{\mathrm{p}}$, which fix a potential difference $\Delta V$ (battery symbol). The water slab (blue) is confined by two Lennard-Jones 9-3 potentials (gray).}

  \label{fig:capacitor}
\end{figure}

\begin{enumerate}
\item[(i)] \textbf{PBC$_{y}$ setup}: Periodic boundary conditions along the $y$ direction were imposed, and the simulation cell was kept square in the $x$–$y$ plane, with lateral dimensions $L_{\parallel}\times L_{\parallel}$. After an equilibration run of $10$ ns, production trajectories of $100$ ns were generated.

\item[(ii)] \textbf{Capacitor setup}:
Figure~\ref{fig:capacitor} shows a schematic representation of the capacitor geometry. Two single-atom-thick gold plates were placed, one at $y=0$ and the other at $y=l_{\mathrm{p}}$, where $l_{\mathrm{p}}$ ranges from $200$ to $1500$~\AA\ (see Appendix~A). Gold atoms were fixed on a triangular lattice with lattice constant $a=2.9416$~\AA. The non-Coulombic interaction between gold atoms and water oxygen atoms was modeled using a Lennard--Jones 12--6 potential with parameters $\sigma_{\mathrm{O-Au}}=3.0585$~\AA\ and $\varepsilon_{\mathrm{O-Au}}=0.9064$~kcal/mol, as in~\cite{electrode}. Simulations under a constant electric field along the $y$ direction were performed using the constant-potential method~\cite{siepmann1995influence,reed2007electrochemical}, as implemented in the \textsc{ELECTRODE} package for LAMMPS~\cite{electrode}, in which the Gaussian charges on the gold atoms are adjusted at every time step so as to maintain a fixed potential difference $\Delta V$ between the plates. The electric field thus arises self-consistently from the electrode and water charges, evaluated by slab-corrected Ewald summation~\cite{yeh1999}. In this finite geometry, $E_{\parallel}=\Delta V/l_{\mathrm{p}}$ denotes the average field between the plates, and the derivative in Eq.~(\ref{def}) is evaluated as $\alpha_{\parallel}^{\mathrm{cap}}=P_{2\mathrm{D}}/(\varepsilon_0 E_{\parallel})$, with $P_{2\mathrm{D}}=\langle M_y\rangle/(L_x l_{\mathrm{p}})$ the average dipole moment along $y$ per unit area of the film. The linearity of the response at $\Delta V=0.5$~V is verified in Appendix~B for $l_{\mathrm{p}}=200$~\AA, the smallest plate separation and hence the largest average field considered. The supercell dimension along $z$ was scaled with the plate separation as $L_z\sim l_{\mathrm{p}}/2$, so as to keep approximately constant the ratio $L_z/l_{\mathrm{p}}$ across the different values of $l_{\mathrm{p}}$ (see Appendix~A).

For large values of $l_{\mathrm{p}}$, the computational cost of the capacitor simulations is significant. To obtain adequate sampling within feasible runtimes, we generate an ensemble of $10$ independent trajectories for each value of $l_{\mathrm{p}}$. Specifically, the confined water was first heated to $600$~K and evolved for $100$~ps. Configurations were then extracted every $25$~ps, a time interval comparable to the dipole-moment relaxation time of the system under these conditions~\cite{zhang2013}, quenched to $300$~K, and used as starting points for independent runs consisting of $1$~ns of equilibration followed by $10$~ns of production. This protocol yields 10 trajectories for each value of $l_{\mathrm{p}}$ initiated from distinct equilibrated configurations, providing structural sampling comparable to that of a single $100$~ns trajectory.
\end{enumerate}

\subsection{Results and discussion}

In this section we estimate $\alpha_{\parallel}$ using both the PBC$_y$ and the capacitor setups.

\subsubsection{PBC$_y$}

\begin{figure}[t]
  \centering
  \includegraphics[width=1.136\columnwidth]{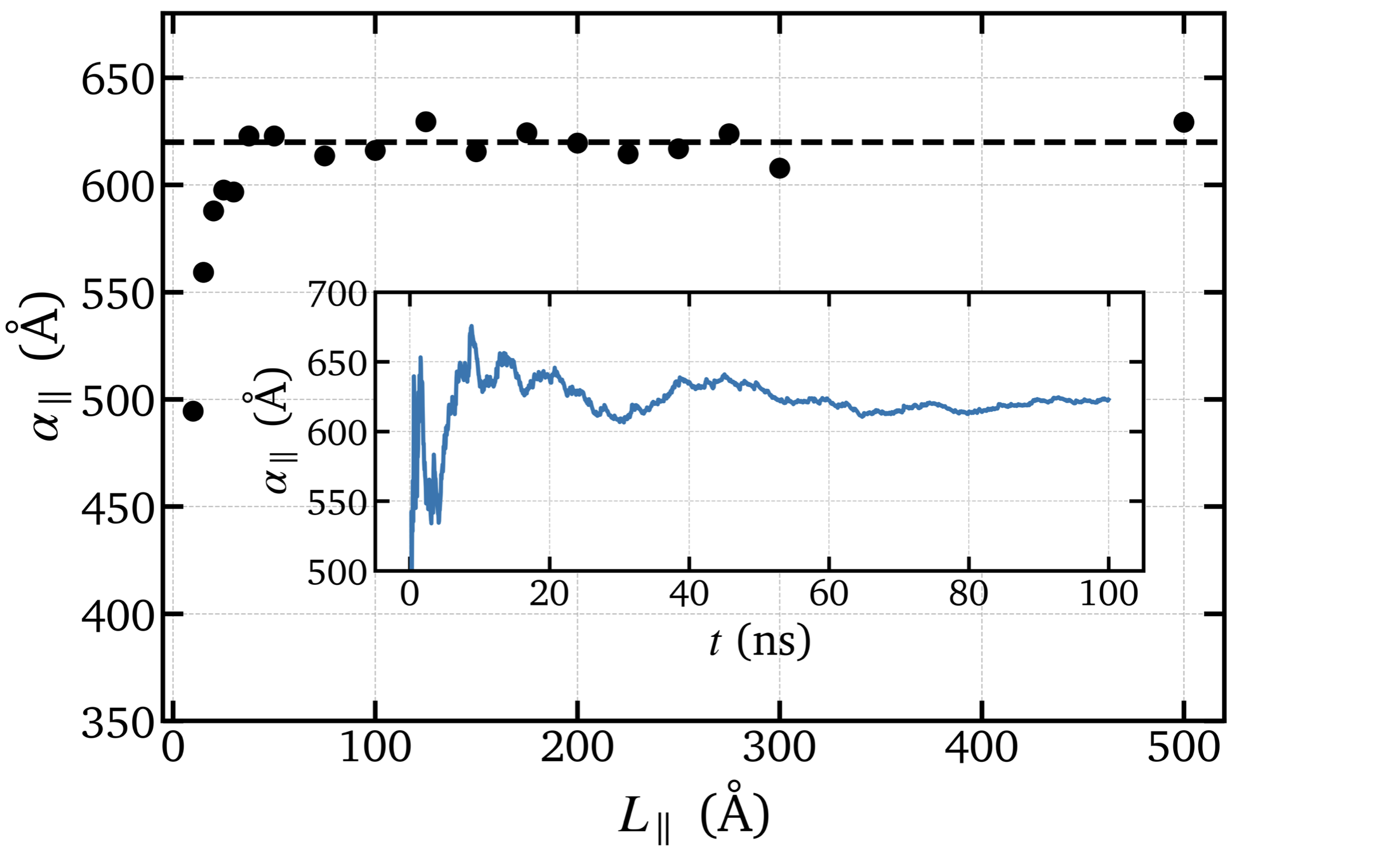}
  \caption{In-plane 2D polarizability $\alpha_{\parallel}$ computed from dipole-moment fluctuations as a function of the lateral supercell size under PBC. The horizontal dashed line indicates the converged value $\alpha_{\parallel}^{\mathrm{PBC}_y}\sim 620$~\AA. Inset: cumulative running average of $\alpha_{\parallel}$ for $L_{\parallel}=37.5$~\AA.}

  \label{fig:convergence}
\end{figure}

We analyze the convergence of the in-plane 2D polarizability with respect to the lateral dimensions of the simulation supercell, $L_{\parallel}$, and the simulation length, as obtained from dipole-moment fluctuations via Eq.~(\ref{eq:alpha_parallel_fluc}). Figure~\ref{fig:convergence} shows the resulting values. We find that convergence is already achieved for $L_{\parallel}=37.5$~\AA, yielding an estimate of $\alpha_{\parallel}^{\mathrm{PBC_{y},fluc}} \sim 620$~\AA\ (Table~I). The inset of Figure~\ref{fig:convergence} shows the cumulative running average of $\alpha_{\parallel}$ for $L_{\parallel}=37.5$~\AA. In line with previous studies~\cite{zhang2013,motevaselian2020}, it shows that long trajectories, on the order of $100$~ns, are required to obtain a converged estimate of the in-plane dipole-moment fluctuations.

\begin{figure*}[t!]
  \centering
  \includegraphics[width=2.21\columnwidth]{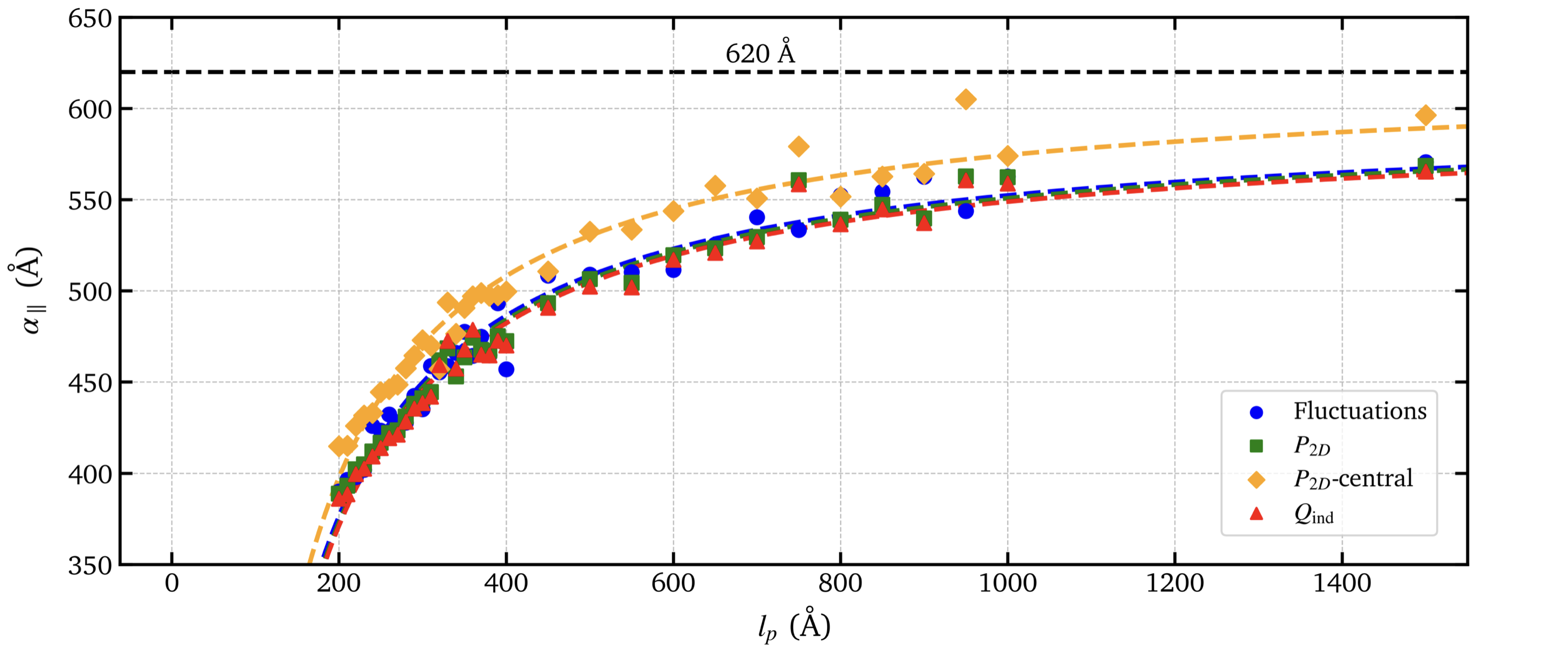}
  \caption{Capacitor-based estimates of the in-plane 2D polarizability, $\alpha_{\parallel}^{\mathrm{cap}}$, as a function of the plate separation $l_{\mathrm{p}}$: fluctuation estimates at $\Delta V=0$ (blue circles), and field-response estimates at $\Delta V=0.5$~V obtained from the induced 2D polarization of the full water slab (green squares) and of its central $100$~\AA\ slice (orange diamonds), together with an estimate inferred from the induced electrode charge (red triangles). Dashed lines are fits to Eq.~(\ref{eq:3}). The horizontal dashed black line shows the PBC$_y$ reference value $\alpha_{\parallel}^{\mathrm{PBC_{y},fluc}}=620$~\AA.}

  \label{fig:alpha}
\end{figure*}

\subsubsection{Capacitor}

In contrast to the PBC$_y$ calculations, the water film in the capacitor setup is finite along $y$, so boundary regions are expected to contribute differently from the central, bulklike part. As a result, the capacitor-based estimate of the in-plane 2D polarizability, $\alpha_{\parallel}^{\mathrm{cap}}$, is expected to be influenced by these edges and to depend on the plate separation $l_{\mathrm{p}}$. Accordingly, $\alpha_{\parallel}^{\mathrm{cap}}$ at finite $l_{\mathrm{p}}$ should be regarded as a geometry-dependent finite-size response. In this finite geometry, the local in-plane electric field is not necessarily uniform, particularly near the film edges. As $l_{\mathrm{p}}$ increases, however, the relative contribution of these edge regions decreases, and the central part of the film approaches the translationally invariant planar limit, in which $E_{\parallel}$ is uniform. The value of $\alpha_{\parallel}$ is therefore obtained in the asymptotic large-$l_{\mathrm{p}}$ limit. To determine this asymptotic value, we model the dependence of $\alpha_{\parallel}^{\mathrm{cap}}$ on $l_{\mathrm{p}}$ as follows. We divide the water layer along $y$ into three domains: two interfacial edge domains next to the gold plates, each extending $l_{\mathrm{e}}$ and characterized by a 2D in-plane polarizability $\alpha_{\parallel,\mathrm{e}}$, and a central bulk-like domain extending $l_{\mathrm{p}}-2l_{\mathrm{e}}$ and 2D polarizability $\alpha_{\parallel}$. Within this model the measured 2D polarizability can be written as
\begin{equation}
\alpha_{\parallel}^{\mathrm{cap}} = \alpha_{\parallel,\mathrm{e}}\frac{2l_{\mathrm{e}}}{l_{\mathrm{p}}}+\alpha_{\parallel}\frac{(l_{\mathrm{p}}-2l_{\mathrm{e}})}{l_{\mathrm{p}}},
\label{eq:2}
\end{equation}
or equivalently as:
\begin{equation}
\alpha_{\parallel}^{\mathrm{cap}} = \alpha_{\parallel}+\frac{2l_{\mathrm{e}}}{l_{\mathrm{p}}}\left(\alpha_{\parallel,\mathrm{e}}-\alpha_{\parallel}\right).
\label{eq:3}
\end{equation}
Equation~(\ref{eq:3}) shows that finite-size corrections decay as $1/l_{\mathrm{p}}$, so that $\alpha_{\parallel}^{\mathrm{cap}}$ approaches $\alpha_{\parallel}$ as the plate separation increases. We therefore compute $\alpha_{\parallel}^{\mathrm{cap}}$ for several values of $l_{\mathrm{p}}$ and fit the resulting data with Eq.~(\ref{eq:3}) by nonlinear least-squares, using \texttt{scipy.optimize.curve\_fit}~\cite{virtanen2020scipy}. The value of $\alpha_{\parallel}$ is then obtained from the fitted curve in the $1/l_{\mathrm{p}}\to 0$ limit. Thus, the finite-$l_{\mathrm{p}}$ values serve as intermediate, geometry-dependent estimates from which the asymptotic response is obtained through the fitting procedure.

Figure~\ref{fig:alpha} shows the capacitor-based estimates of $\alpha_{\parallel}^{\mathrm{cap}}$ obtained using several methods for different values of $l_{\mathrm{p}}$. The blue symbols correspond to simulations at zero applied bias ($\Delta V=0$) and to $\alpha_{\parallel}^{\mathrm{cap}}$ computed from dipole-moment fluctuations via Eq.~(\ref{eq:alpha_parallel_fluc}). Note that Eq.~(\ref{eq:alpha_parallel_fluc}) is derived for the translationally invariant periodic system and not for the capacitor geometry, where the explicit electrodes break the in-plane invariance. It is used here only as an operational estimator whose large-$l_{\mathrm{p}}$ limit is expected to recover the periodic result. Fitting these data with Eq.~(\ref{eq:3}) (dashed blue line) yields $\alpha_{\parallel}^{\mathrm{cap,fluc}}=596.3$~\AA\ (Table~I).

\begin{table}[t]
\centering
\caption{\label{tab:mytable}
In-plane 2D polarizability $\alpha_{\parallel}$ obtained using the different methods considered in this work. $\alpha_{\parallel}^{\mathrm{PBC_{y},fluc}}$ is computed in the PBC$_{y}$ setup from dipole-moment fluctuations [Eq.~(\ref{eq:alpha_parallel_fluc})]. The remaining estimates are obtained in the capacitor setup: $\alpha_{\parallel}^{\mathrm{cap,fluc}}$ from dipole-moment fluctuations at $\Delta V=0$ [Eq.~(\ref{eq:alpha_parallel_fluc})], $\alpha_{\parallel}^{\mathrm{cap},P_{2\mathrm{D}}}$ from the induced 2D polarization at $\Delta V=0.5$~V [Eq.~(\ref{def})], $\alpha_{\parallel}^{\mathrm{cap},P_{2\mathrm{D}}\text{-central}}$ computed as $\alpha_{\parallel}^{\mathrm{cap},P_{2\mathrm{D}}}$ but restricting the analysis to the central region of the water slab (see main text), and $\alpha_{\parallel}^{\mathrm{cap},Q_{\mathrm{ind}}}$ obtained from the induced charge on the capacitor plates [see Eq.~(\ref{eq:9})]. The uncertainties correspond to standard errors. All values are reported in \AA.}
\begin{ruledtabular}
\begin{tabular}{ccccc}
$\alpha_{\parallel}^{\mathrm{PBC_{y},fluc}}$ &
$\alpha_{\parallel}^{\mathrm{cap,fluc}}$ &
$\alpha_{\parallel}^{\mathrm{cap},P_{2\mathrm{D}}}$ &
$\alpha_{\parallel}^{\mathrm{cap},P_{2\mathrm{D}}\text{-central}}$ &
$\alpha_{\parallel}^{\mathrm{cap},Q_{\mathrm{ind}}}$ \\
\noalign{\vskip 2.5pt}
\hline
\noalign{\vskip 2.5pt}
$620 \pm 4$ &
$596 \pm 4$ &
$595 \pm 4$ &
$619 \pm 4$ &
$593 \pm 4$ \\
\end{tabular}
\end{ruledtabular}
\end{table}

The green symbols are obtained from the induced 2D polarization (Sec.~III.A). Fitting these data with Eq.~(\ref{eq:3}) (dashed green line) gives $\alpha_{\parallel}^{\mathrm{cap},P_{2D}}=595.2$~\AA\ (Table~I). These two estimates are in excellent agreement despite being obtained from independent simulations. Both values, however, are slightly smaller than that obtained in the PBC$_y$ setup. In particular, the $\Delta V=0$ capacitor estimates are expected to recover $\alpha_{\parallel}^{\mathrm{PBC_{y},fluc}}$ once $l_{\mathrm{p}}$ is sufficiently large~\cite{zhang2016computing}. The fact that the two do not yet coincide indicates that, over the range of $l_{\mathrm{p}}$ explored here, the edge regions still contribute measurably to the estimate of the 2D polarizability.

To further test whether this discrepancy originates from boundary effects, we recomputed the 2D polarization under $\Delta V=0.5$~V by restricting the analysis to the central $100$~\AA\ slice of the water film, thereby reducing the contribution of the edge regions. The size of the central region was chosen to balance suppressing edge contributions against retaining a sufficiently large analysis region for every plate separation in Fig.~\ref{fig:alpha}. Since the smallest capacitor has $l_{\mathrm{p}}=200$~\AA, a $100$~\AA\ slice retains all data points while excluding a substantial fraction of the edge-affected region and preserving enough central area for a reliable polarization estimate. As expected, the resulting values (orange symbols in Figure~\ref{fig:alpha}) are systematically larger than those obtained from the 2D polarization of the full slab. Fitting the orange data with Eq.~(\ref{eq:3}) (dashed orange line) yields $\alpha_{\parallel}^{\mathrm{cap},P_{2D}\text{-central}}=618.6$~\AA\ (Table~I), in excellent agreement with $\alpha_{\parallel}^{\mathrm{PBC_{y},fluc}}$.

Overall, we find excellent agreement between the PBC and capacitor-based estimates of $\alpha_{\parallel}$, showing the internal consistency of the two approaches. This agreement provides a numerical verification that the large-$l_{\mathrm{p}}$ limit of the capacitor setup recovers the translationally invariant planar response. From a practical point of view, however, the PBC route is vastly more computationally efficient. In the capacitor geometry, boundary effects associated with the finite extent of the water slab along $y$ remain appreciable even at our largest value of $l_{\mathrm{p}}$. Moreover, these simulations are extremely costly: for $l_{\mathrm{p}} = 1500$~\AA, generating $\sim 11$~ns of dynamics in the capacitor geometry required about $26$ days of runtime on $64$ CPU cores. By contrast, in the PBC$_y$ setup with $L_{\parallel}=37.5$~\AA, a $100$~ns trajectory can be produced in about $1.5$ days on $32$ CPU cores. Comparing the total computational cost (runtime $\times$ number of cores), the PBC-based fluctuation method is therefore at least on the order of $300$ times more efficient than the capacitor approach for obtaining an estimate of $\alpha_{\parallel}$ with comparable reliability.

The converged value, $\alpha_{\parallel}\sim 620$~\AA, can be interpreted in terms of a large in-plane screening radius. In the sense discussed by Tian \textit{et al.}~\cite{tian2019}, this corresponds to a large in-plane length scale governing the gradual crossover from the short-distance screened interaction to the asymptotic vacuum Coulomb behavior. By comparison, the corresponding characteristic length scale in the perpendicular direction is much smaller, of the order of $6$~\AA~\cite{zubeltzu2025}, indicating a marked dielectric anisotropy under confinement.

We now address a complementary question of practical relevance: whether $\alpha_{\parallel}$ can be inferred directly from electrode charging, as in conventional capacitor measurements. In electrostatics, in an ideal three-dimensional parallel-plate geometry, placing a dielectric between two plates held at fixed voltage causes the dielectric to polarize and induces a compensating surface charge on the electrodes, where the induced surface charge density coincides with the polarization of the dielectric. By direct dimensional analogy, in our capacitor setup, a finite 2D dielectric film is expected to induce a line charge density on the electrodes localized in front of the film. This induced line charge density is expected to coincide with the film’s 2D polarization:
\begin{equation}
\lambda_{\mathrm{ind}} = P_{2\mathrm{D}},
\label{eq:lambda}
\end{equation}
where $\lambda_{\mathrm{ind}}$ is the induced charge per unit length along the edge of the 2D dielectric, in our case, along the $x$ direction (see Figure~\ref{fig:capacitor}). This relation provides an alternative way to estimate $\alpha_{\parallel}$ by substituting $P_{2\mathrm{D}}$ with $\lambda_{\mathrm{ind}}$ in Eq.~(\ref{def}).

\begin{figure}[t]
  \centering
  \includegraphics[width=1.10\columnwidth]{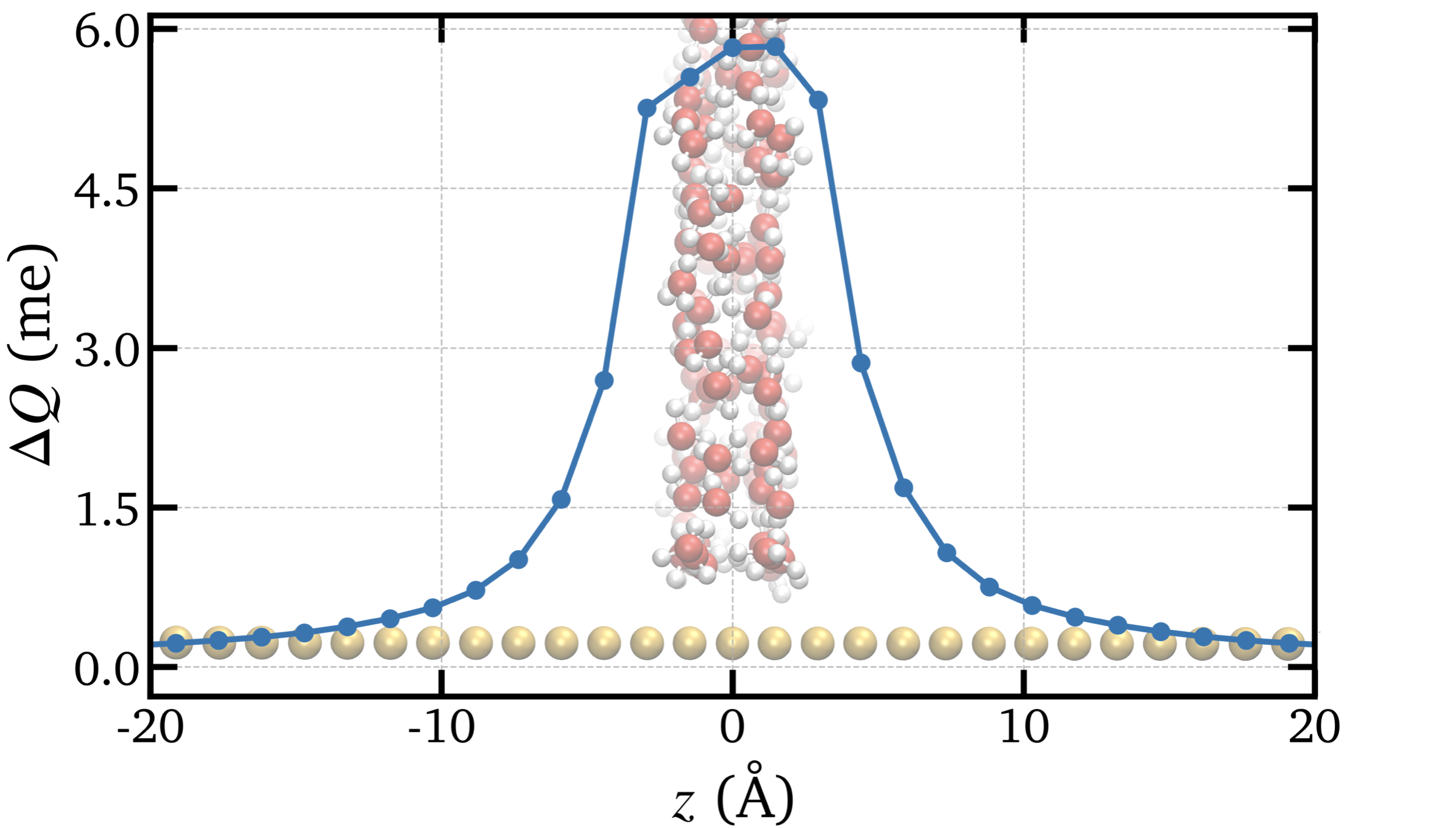}
  \caption{Charge difference $\Delta Q(z_i)$ on the gold-atom rows of the positive electrode between simulations at $\Delta V=0.5$~V and $\Delta V=0$~V, as a function of the $z$ coordinate for $l_{\mathrm{p}}=950$~\AA\ (blue line with circles). The background snapshot shows the electrode rows (gold spheres) and the confined water slab, with oxygen atoms in red and hydrogen atoms in white.}

  \label{fig:charge}
\end{figure}

Figure~\ref{fig:charge} shows, for the capacitor setup with $l_{\mathrm{p}}=950$~\AA, the charge difference $\Delta Q(z_i)$ on the positive electrode between simulations at $\Delta V=0.5$~V and at $\Delta V=0$~V. Specifically, $\Delta Q(z_i)$ is computed by summing the charge differences of the gold atoms belonging to the same electrode row, i.e. atoms that share the same coordinate $z_i$, with $i=1,\dots,N_{\mathrm{rows}}$, where $N_{\mathrm{rows}}$ is the total number of gold-atom rows along the $z$ direction. For visual guidance, a snapshot of part of the simulation cell is shown in the background of Figure~\ref{fig:charge}: the electrode atoms are arranged in rows shown as a sequence of spheres, and each value of $z_i$ in the plot corresponds to one such row of gold atoms. The charge difference is localized in the range of $z$ where the water is located, forming a rod like profile with a characteristic thickness of $\sim 2$~nm that is comparable to the extent of the confined water region. This spatial location of $\Delta Q(z_i)$ in front of the water film is consistent with the dimensional argument above, namely that a 2D dielectric produces a one-dimensional induced charge distribution on the electrodes.

To obtain the induced charge $Q_{\mathrm{ind}}$ in the positive electrode, we subtract the electrode charge measured in simulations without water, $Q_0$, from that measured in otherwise identical simulations with the water slab present,
\begin{equation}
Q_{\mathrm{ind}}=\sum_{i=1}^{N_{\mathrm{rows}}}\Delta Q(z_i)-Q_0.
\end{equation}
As mentioned above, combining Eq.~(\ref{def}) with Eq.~(\ref{eq:lambda}) allows us to test whether the in-plane 2D polarizability can be obtained directly from electrode charging. The resulting values (red symbols in Figure~\ref{fig:alpha}) are in excellent agreement with $\alpha_{\parallel}$ obtained from the induced 2D polarization of the water slab under bias (green symbols in Figure~\ref{fig:alpha}), showing that $\alpha_{\parallel}$ can be accessed through electrode charging. These results further indicate that, as for the perpendicular 2D polarizability $\alpha_{\perp}$~\cite{zubeltzu2025}, $\alpha_{\parallel}$ can, in principle, be obtained experimentally by capacitance measurements. In the idealized large plate separation limit, where edge effects are negligible, $\alpha_{\parallel}$ can be inferred from capacitance measurements obtained with ($C$) and without ($C_0$) water as
\begin{equation}
\alpha_{\parallel}=\frac{l_{\mathrm{p}}(C-C_0)}{L_x\varepsilon_0}.
\label{eq:9}
\end{equation}

\subsubsection{Sensitivity of $\varepsilon_{\parallel}$ to width}

\begin{figure}[t]
  \centering
  \includegraphics[width=1.1\columnwidth]{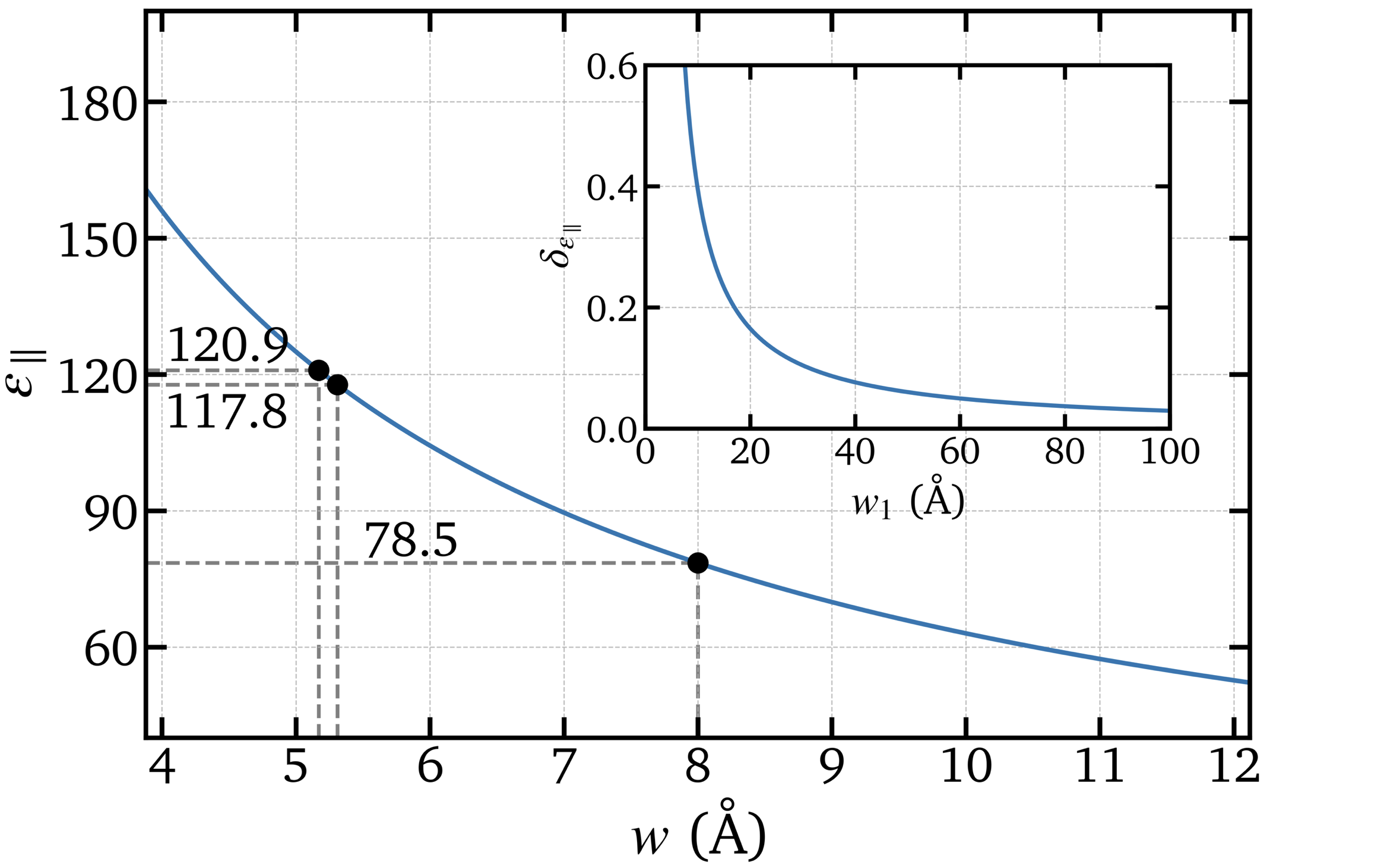}
  \caption{In-plane relative permittivity $\varepsilon_{\parallel}$ computed from $\alpha_{\parallel}=620$~\AA\ as a function of the chosen film width $w$ (blue solid line). Vertical dashed lines indicate three choices of $w$ used in the literature (see main text). The inset reports the relative difference $\delta_{\varepsilon_{\parallel}}$ (see main text) as a function of $w_1$.}

  \label{fig:epsilon}
\end{figure}

We now analyze how the choice of an effective film thickness affects the in-plane dielectric constant. Figure~\ref{fig:epsilon} illustrates the sensitivity of $\varepsilon_{\parallel}$ to the chosen width $w$, using $\alpha_{\parallel}=620$~\AA\ as obtained in this work. Dashed vertical lines indicate three choices of $w$ that have been employed in the literature~\cite{zubeltzu2025}: (1) the separation between the asymptotes of the confining potentials, $w_1=8$~\AA; (2) $w_2=5.17$~\AA, the accessible perpendicular space as defined by Kumar \textit{et al.}~\cite{kumar2005}; and (3) $w_3=5.31$~\AA, obtained by normalizing the film’s 2D molecular density by the bulk 3D density of water. Notably, the first choice, which is the one most commonly used in practice, yields $\varepsilon_{\parallel} = 78.5$, whereas the latter two definitions give values that are about $50\%$ larger (see Figure~\ref{fig:epsilon}).

In addition, we can estimate how the convention dependence of the in-plane permittivity evolves as the confining width is increased. A relevant and commonly encountered limit is $\alpha_{\parallel}\gg w$: it holds in the present few-layer regime (here $\alpha_{\parallel}\sim 620$~\AA\ while $w$ is of order $1$~nm), it is also consistent with simulations and experiments reporting very large in-plane permittivities under extreme confinement~\cite{motevaselian2020,renou2015,hamid2021,Wang2025}, and it even applies in the bulk limit, where $\varepsilon\sim 80$. In this limit, Eq.~\eqref{definition_alpha} reduces to $\varepsilon_{\parallel} = \alpha_{\parallel}/w$, and therefore, using two width conventions gives permittivities whose relative difference is controlled by the width ratio. To obtain an upper-bound estimate within the set of definitions considered above, we take $w_1$ and $w_2$ to be the two choices given that they differ the most. Defining the relative difference with respect to $\varepsilon_{\parallel}(w_1)$ as
\begin{equation}
\delta_{\varepsilon_{\parallel}}
\equiv
\frac{\varepsilon_{\parallel}(w_2)-\varepsilon_{\parallel}(w_1)}{\varepsilon_{\parallel}(w_1)},
\end{equation}
it follows directly that
\begin{equation}
\delta_{\varepsilon_{\parallel}}=\frac{w_1}{w_2}-1.
\end{equation}
For the width definition of Kumar \textit{et al.}~\cite{kumar2005}, one has in our case the simple relation $w_2=w_1-2.83$~\AA, so that $\delta_{\varepsilon_{\parallel}}$ is determined entirely by $w_1$. The inset of Fig.~\ref{fig:epsilon} shows $\delta_{\varepsilon_{\parallel}}$ as a function of width $w_1$. For example, a relative difference of $\delta_{\varepsilon_{\parallel}}= 0.05$ is obtained at $w_1\sim 60$~\AA, indicating that the convention dependence associated with the width choice drops to a few percent once the confinement width approaches the $\sim 10$~nm scale. For nanometer-scale slit pores, $\varepsilon_{\parallel}$ thus depends strongly on the convention adopted for $w$, as expected from Eq.~(\ref{definition_alpha}), whereas the integrated response $\alpha_{\parallel}$ is independent of this convention.

\section{Out-of-plane 2D polarizability}

We also verify that $\alpha_{\perp}$ can be obtained from Eq.~(\ref{eq:alpha_perp_fluc}) through the fluctuation-dissipation theorem. To do so, we adopt the same computational setup as in Ref.~\cite{zubeltzu2025}, using the TIP4P/2005 water model~\cite{abascal2005general}. The only difference is that we increase the production length from $10$~ns to $100$~ns in order to remain consistent with the simulations reported in this work. We perform two types of calculations: (i) simulations without an applied perpendicular electric field, from which $\alpha_{\perp}$ is estimated via Eq.~(\ref{eq:alpha_perp_fluc}); and (ii) simulations with an applied perpendicular field, from which $\alpha_{\perp}$ is obtained from the induced perpendicular 2D polarization, following exactly the procedure of Ref.~\cite{zubeltzu2025}. Both approaches correspond to a constant displacement-field ensemble along the confinement direction.

\begin{figure}[t]
  \centering
  \includegraphics[width=1.10\columnwidth]{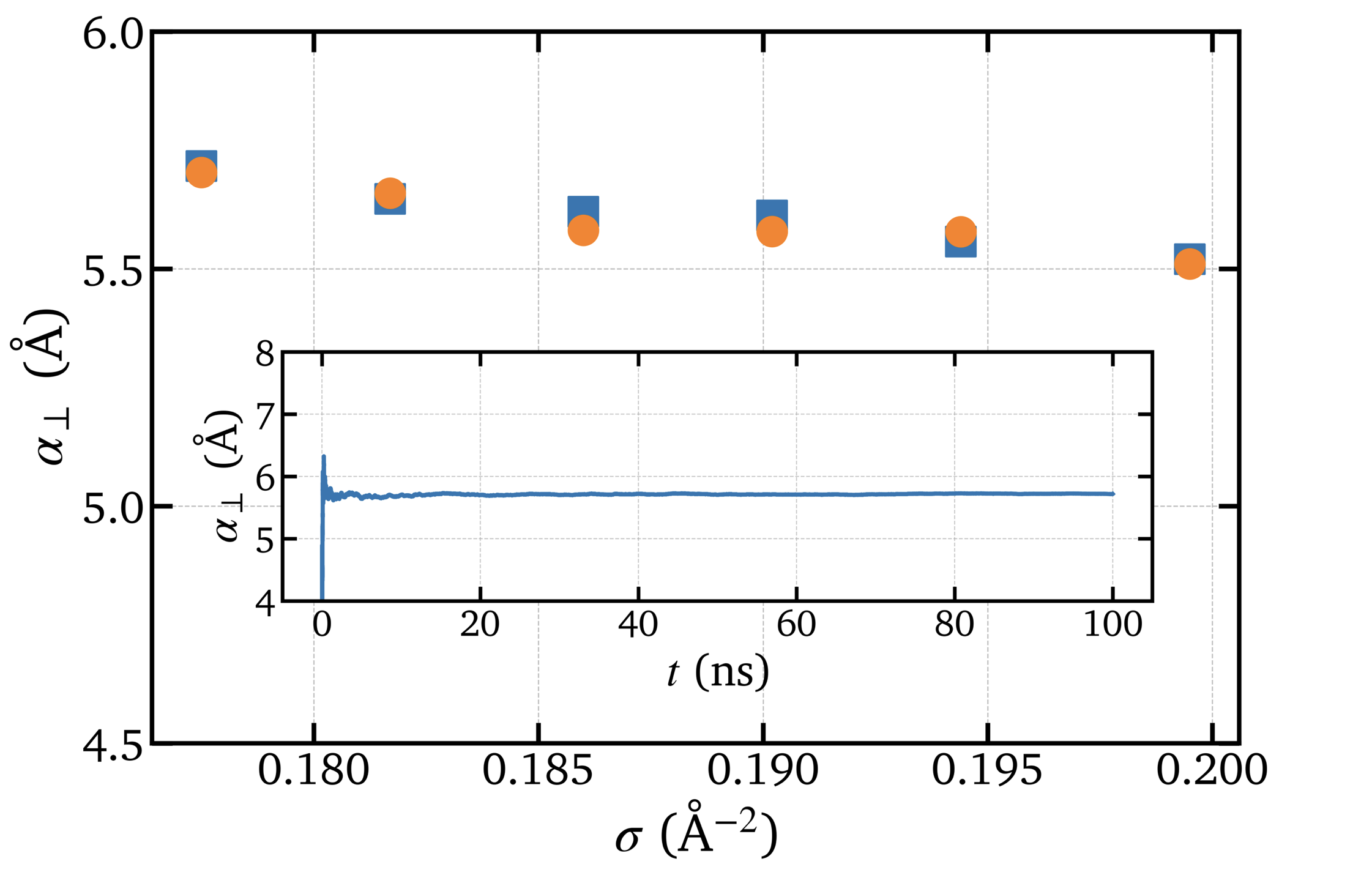}
  \caption{Perpendicular 2D polarizability, $\alpha_{\perp}$, as a function of the 2D molecular density $\sigma$ for TIP4P/2005 water. Values obtained from equilibrium dipole-moment fluctuations using Eq.~(\ref{eq:alpha_perp_fluc}) (blue squares) are consistent with those extracted from the induced perpendicular 2D polarization under an applied perpendicular field following Ref.~\cite{zubeltzu2025} (orange circles). Inset: Cumulative running average of $\alpha_{\perp}$ at $\sigma=0.177$~\AA$^{-2}$ obtained with Eq.~(\ref{eq:alpha_perp_fluc}).}

  \label{fig:perpendicular}
\end{figure}

Figure~\ref{fig:perpendicular} shows the resulting $\alpha_{\perp}$ as a function of the 2D molecular density $\sigma$, over the same range considered in Ref.~\cite{zubeltzu2025}. The two routes yield excellent agreement, supporting the robustness of the estimate. The inset of Figure~\ref{fig:perpendicular} shows the cumulative running average of $\alpha_{\perp}$ at $\sigma=0.177$~\AA$^{-2}$, obtained in the absence of an applied voltage using the fluctuation-dissipation expression [Eq.~(\ref{eq:alpha_perp_fluc})]. Consistent with previous observations that the dipole moment relaxes much faster along the out-of-plane direction in nanometer slits than in the in-plane directions~\cite{zhang2013}, we find that $\alpha_{\perp}$ converges rapidly, and trajectories of $\sim 10$~ns are sufficient to obtain a converged value.

Within the PBC$_y$ setup used in this work, Eq.~(\ref{eq:alpha_perp_fluc}) also allows us to estimate $\alpha_{\perp}$ for SPC/E water at $\sigma=0.177$~\AA$^{-2}$. We obtain $\alpha_{\perp}=6.0$~\AA, which is close to the value found above for TIP4P/2005 at the same 2D molecular density, $\alpha_{\perp}=5.7$~\AA. Together with the corresponding in-plane value $\alpha_{\parallel}\sim 620$~\AA, this allows us to quantify the dielectric anisotropy in the spirit of Tian \textit{et al.}~\cite{tian2019} through the anisotropy index $\eta=\alpha_{\perp}/\alpha_{\parallel}$. For the present system, this yields $\eta\sim10^{-2}$, evidencing the strongly anisotropic dielectric response of nanoconfined water.

\section{Conclusion}

In this work we introduce the in-plane 2D polarizability $\alpha_{\parallel}$ as an appropriate descriptor of in-plane screening in the few-layer regime. Using classical MD simulations of SPC/E water confined in an $\sim 8$~\AA\ slit, we computed $\alpha_{\parallel}$ using two independent and mutually consistent approaches. First, in the PBC$_y$ setup, $\alpha_{\parallel}$ is obtained directly from equilibrium dipole-moment fluctuations through a 2D fluctuation-dissipation relation. Second, in the capacitor-based setup, $\alpha_{\parallel}$ is estimated at zero applied voltage from polarization fluctuations and is also obtained from the 2D polarization under an applied in-plane field. All these approaches yield consistent results, with $\alpha_{\parallel}\sim 620$~\AA\ for the conditions studied here. This large value indicates that the confined water layer influences in-plane electrostatic interactions over remarkably long distances, reflecting a large in-plane screening radius and, therefore, a large in-plane length scale governing the gradual crossover from the short-distance screened interaction to the asymptotic vacuum Coulomb behavior.

The capacitor-based simulations also show how $\alpha_{\parallel}$ can be accessed experimentally. In this geometry, the charge induced on the electrodes and localized in front of the water film provides a direct measure of the 2D polarization, allowing $\alpha_{\parallel}$ to be determined from electrode charging and capacitance measurements. This establishes a concrete way of comparing simulations and experiments in terms of a 2D response function.

As an additional validation of the fluctuation-based approach, we verified that the perpendicular 2D polarizability, $\alpha_{\perp}$, obtained from equilibrium fluctuations is in agreement with the established field-response results reported for TIP4P/2005 water over a range of 2D densities~\cite{zubeltzu2025}. This supports the robustness of the 2D formulation for both components of the dielectric response. Together with the much smaller characteristic length scale previously reported for the perpendicular direction, the present results confirm the strong dielectric anisotropy of the confined film. They also show that this anisotropy can be quantified directly in terms of the integrated 2D response, without the need for an effective thickness.

While the specific confinement conditions considered here do not reproduce the giant in-plane permittivities reported in~\cite{Wang2025}, this is not unexpected, since the microscopic state of the confined water in those experiments, as well as the corresponding confinement conditions, cannot be established straightforwardly \emph{a priori}. At the same time, simulations using the SPC/E water model, as in the present work, have reported in-plane responses of similar magnitude when the confinement accommodates a single water layer~\cite{motevaselian2020}. The emphasis of the present work is therefore not on matching a particular reported value, but on establishing a practical and internally consistent protocol, based on the 2D polarizability, for quantifying the in-plane dielectric response at molecular scales. The framework introduced here, combined with that established previously for the perpendicular response in Ref.~\cite{zubeltzu2025}, is general and can be extended to other polar liquids, thereby providing a systematic basis to reassess and compare dielectric properties in both the in-plane and perpendicular directions without relying on width conventions.

Although water is used here as a case study, the proposed formulation follows from linear-response theory and is expected to apply generally to polar fluids under quasi-2D confinement, provided that the response remains linear and the appropriate electrostatic boundary conditions are used. A systematic assessment of the framework for other confined polar liquids is left for future work.

\begin{acknowledgments}
We acknowledge financial support from the Basque Government (Eusko Jaurlaritza) through Project No. IT1584-22. We also acknowledge the SGIker/IZO-SGIker facilities of the University of the Basque Country (UPV/EHU), supported by MICINN, GV/EJ, ERDF and ESF, as well as the DIPC Supercomputing Center, for generous allocations of computational resources and technical support. DIPC was supported by the ``Severo Ochoa'' Programme for Centres of Excellence in R\&D (MINECO, CEX2018-000867-S). We are especially grateful to Dr.~Óscar Pozo for his invaluable support and guidance throughout this project. His insightful ideas and suggestions played a key role in shaping the direction of this work, and it benefited greatly from his advice. We also thank Prof.~Emilio Artacho and Dr.~Jon Romero for useful discussions, and Luis Paniego and Iker Videira for supporting and carrying out the design of Fig.~1.
\end{acknowledgments}

\section*{Data availability}
The simulation input files and analysis data that support the findings of this article are openly available at Ref.~\cite{alpha_parallel_repo}.

\appendix

\section{Capacitor-plates dimensions}

Table~\ref{tab:dimensions} shows the dimension $L_z$ of the supercell along the $z$ direction employed for the capacitor-based calculations. For each plate separation $l_{\mathrm{p}}$ we chose an $L_z \sim l_{\mathrm{p}}/2$, so as to keep the ratio $L_z/l_{\mathrm{p}}$ approximately constant across the explored range. The number of water molecules $N$ is adjusted to maintain the fixed 2D molecular density $\sigma=0.177$~\AA$^{-2}$ at each value of $l_{\mathrm{p}}$.

\begin{table}[h]
\centering
\caption{\label{tab:dimensions}Simulation-cell parameters used in the capacitor setup: plate separation $l_{\mathrm{p}}$, supercell length $L_z$, and number of water molecules $N$ at fixed 2D molecular density $\sigma=0.177$~\AA$^{-2}$.}
\begin{ruledtabular}
\begin{tabular}{ccc}
$l_{\mathrm{p}}$ (\AA) & $L_z$ (\AA) & $N$ \\
\hline
200 & 117.663 & 1187 \\
210 & 117.663 & 1260 \\
220 & 117.663 & 1306 \\
230 & 117.663 & 1365 \\
240 & 141.1956 & 1424 \\
250 & 141.1956 & 1484 \\
260 & 141.1956 & 1543 \\
270 & 141.1956 & 1602 \\
280 & 164.7282 & 1662 \\
290 & 164.7282 & 1721 \\
300 & 164.7282 & 1781 \\
310 & 164.7282 & 1840 \\
320 & 164.7282 & 1899 \\
330 & 188.2608 & 1959 \\
340 & 188.2608 & 2018 \\
350 & 188.2608 & 2077 \\
360 & 188.2608 & 2137 \\
370 & 211.7934 & 2196 \\
380 & 211.7934 & 2255 \\
390 & 211.7934 & 2315 \\
400 & 211.7934 & 2374 \\
450 & 235.326 & 2670 \\
500 & 282.3912 & 2967 \\
550 & 305.9238 & 3264 \\
600 & 329.4564 & 3561 \\
650 & 352.989 & 3858 \\
700 & 376.5216 & 4155 \\
750 & 400.0542 & 4451 \\
800 & 447.1194 & 4748 \\
850 & 470.652 & 5045 \\
900 & 494.1846 & 5342 \\
950 & 517.7172 & 5638 \\
1000 & 541.2498 & 5934 \\
1500 & 823.641 & 8901 \\
\end{tabular}
\end{ruledtabular}
\end{table}

\section{Linear-response check in the capacitor setup}

\begin{figure}[h]
  \centering
  \includegraphics[width=1.02\columnwidth]{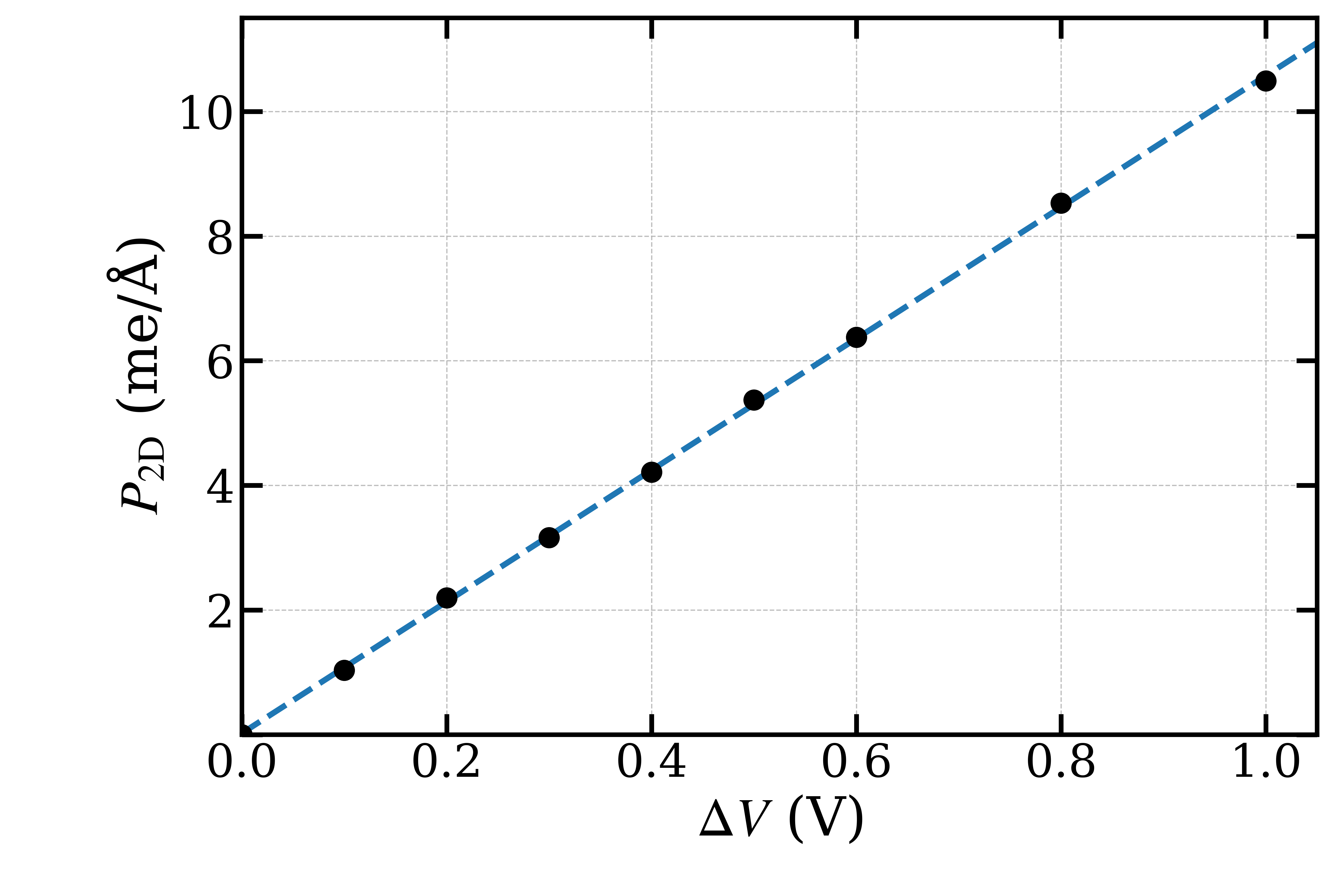}
  \caption{Induced 2D polarization $P_{2\mathrm{D}}$ as a function of the applied potential difference $\Delta V$ in the capacitor setup with $l_{\mathrm{p}}=200$~\AA. The dashed line is a linear fit.}
  \label{fig:linearity}
\end{figure}

Equation~(\ref{def}) assumes linear response. To verify that the applied potential difference $\Delta V=0.5$~V lies within the linear regime, we computed the induced 2D polarization for $\Delta V$ between $0.1$ and $1.0$~V at $l_{\mathrm{p}}=200$~\AA, the smallest plate separation considered in this work and hence the one with the largest average field $E_{\parallel}=\Delta V/l_{\mathrm{p}}$ for a given $\Delta V$. Figure~\ref{fig:linearity} shows that $P_{2\mathrm{D}}$ is linear in $\Delta V$ over the whole range ($R^2=0.9997$), with an intercept compatible with zero ($0.02\pm0.03$~me/\AA) and $\Delta V=0.5$~V in the middle of the range. The slope, $10.55\pm0.06$~me/(\AA\,V), corresponds to $\alpha_{\parallel}^{\mathrm{cap}}=382\pm2$~\AA, consistent with the value obtained at $\Delta V=0.5$~V in Fig.~\ref{fig:alpha}. Linearity at $\Delta V=0.5$~V therefore holds for all plate separations considered.

\clearpage

\bibliography{main}

\end{document}